\documentclass[aps,prl,twocolumn,showpacs,groupedaddress]{revtex4-1}
\usepackage[dvips]{graphicx}
\begin{document}

\title{Proposal for a Self-Excited Electrically Driven\\
Surface Plasmon Polariton Generator}

\author{V.G.~Bordo}
\email{bordo@mci.sdu.dk}

\affiliation{NanoSyd, Mads Clausen Institute, Syddansk Universitet, Alsion 2, DK-6400 S{\o}nderborg, Denmark}


\date{\today}

\begin{abstract}
We propose a generator of surface plasmon polaritons (SPPs) which, unlike spasers or plasmon lasers, does not require stimulated emission in the system. Its principle of operation is based on a positive feedback which a classical oscillating dipole experiences from a reflective surface located in its near field. The generator design includes a nanocavity between two metal surfaces which contains metal nanoparticles in its interior. The whole structure is placed onto a prism surface that allows one to detect the generated SPPs in the Kretschmann configuration. The generation process is driven by a moderate constant voltage applied between the metal covers of the cavity. Both the generation criterion and the steady-state operation of the generator are discussed.
\end{abstract}


\maketitle

\begin{figure}
\includegraphics[width=\linewidth]{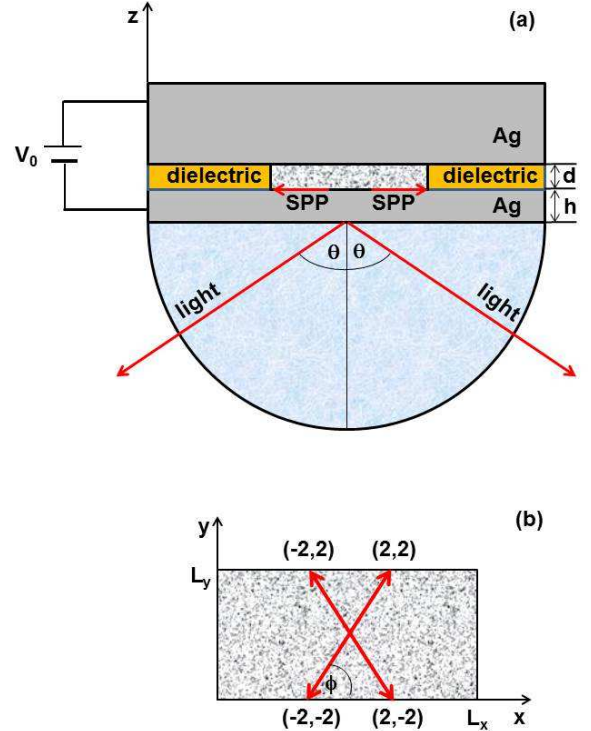}
\caption{\label{fig:sketch} (a) Design of the SPP generator (side view). The leakage radiation of the generated SPPs can be observed in the Kretschmann configuration. (b) Top view of the cavity. The red arrows show the propagation directions of the generated modes $(m,n)$.}
\end{figure}

\emph{Introduction}. -- The progress of nanophotonics develops towards smaller photonic elements and higher packing density of photonic circuitries. Metal nanostructures, which support strongly localized electromagnetic excitations known as surface plasmon polaritons (SPPs), provide unique opportunities in this direction  \cite{Atwater07,Ebbesen08}. Utilization of SPPs allows one to overcome the diffraction limit in photonics \cite{Barnes03} and can be implemented for extreme light energy concentration \cite{Schuller10}, ultra-sensitive sensing \cite{Zayats09}, high-resolution microscopy \cite{Novotny03}, ultra-fast computations and a plenty of other applications.\\
This strategy faces, however, challenges because of high dissipation which is inevitable for all metallic structures. To compensate optical losses, it was suggested to introduce a gain medium into a metal nanostructure \cite{Sudarkin89}. This idea stems from the conventional approach in laser physics \cite{Pantell69}. An active (gain) medium contains atoms or molecules which are transferred, under optical pumping, into their excited states. The excited state population determines the rate of the emission stimulated by the electromagnetic field in the cavity, while the ground state population dictates the stimulated absorption rate. If the population of the excited atoms exceeds the population of the ground-state atoms, i.e. a population inversion occurs, the emission prevails over the absorption. If, besides that, this imbalance overcomes the loss rate for the cavity field, the field is amplified after each round trip in the laser cavity, that represents a loop gain. As a result, the field intensity steadily increases with time until the saturation comes into play. \\
The process described above is an essentially quantum phenomenon. The stimulated emission is accompanied by the atom transfer to the ground state while the released energy goes into the creation of a field quantum (photon). The same principle forms the basis of Surface Plasmon Amplification by Stimulated Emission of Radiation (SPASER) \cite{Stockman03,Stockman11} or plasmon laser \cite{Ma13}. In such a case, the stimulated emission in the gain medium incorporated into the metal nanostructure leads to generation of SPP quanta.\\
This mechanism is, however, not the only possible way of generating SPPs. A classical dipole oscillating above a conductive surface excites SPPs at the surface - the result which dates back to Sommerfeld's paper from 1909 \cite{Sommerfeld09,Sommerfeld64}. A surface can be regarded as an open cavity where the electromagnetic field radiated by the dipole is reflected back to it. The forward and backward pathways interfere with each other and the resulting field at the dipole position depends on the dipole-surface distance, $h$. Sommerfeld calculated the power needed by the dipole to compensate the overall losses which are represented by the radiation into the half-plane above the dipole position and the Joule heat in the underlying conductive substrate. This quantity demonstrates an oscillating behavior as a function of $h/\lambda$, with $\lambda$ being the wavelength of the radiation, that reflects constructive and destructive interference of the radiated and reflected fields.\\
A remarkable result takes place for a horizontal dipole above an infinitely conductive substrate: For very short distances such that $h\ll\lambda$ the power needed by the dipole is equal to zero. In other words, a reflective surface provides a feedback which supports the dipole oscillations and can compensate their decay. The abandonment of the assumption of an infinite substrate conductivity destroys complete loss compensation, however, the relaxation rate of the dipole oscillations remains significantly reduced \cite{Sommerfeld09,Sommerfeld64}.\\
Let us turn now to an ensemble of dipoles oscillating above the surface. Then each individual dipole undergoes the action of the fields of all other dipoles. The direct dipole-dipole radiative interaction is nothing else than the contribution to the Lorentz local field, whereas their interaction via the reflected fields provides an additional feed for the dipole oscillations. If the phases of the latter fields are such that they support oscillations (i.e. they supply a positive feedback), then, for a large enough number of dipoles, the feed can exceed the relaxation. In such a case, the polarization of the ensemble will increase after each radiation-reflection cycle, thus indicating a loop gain. \\
This scenario can be realized for an ensemble of metal nanoparticles (NPs) embedded into a cavity with metallic walls \cite{Bordo16}. Within a certain range of parameters, the field in such a structure can be unstable that leads to its self-excitation (self-oscillation) in the presence of an external resonant field. However self-oscillation can, in principle, be stimulated by a source of power that is not related anyhow with the periodicity of the sustained oscillations, in particular, by a \emph{constant} external field \cite{Jenkins13}. In the present paper, we propose a novel principle of SPP generation in a metal nanostructure which is driven by a constant applied voltage.\\
\emph{System}. -- The design of such a generator is shown in Fig. \ref{fig:sketch}. A rectangular cavity of cross-section $L_x\times L_y$ in the $xy$ plane is formed in the subwavelength gap of thickness $d$ between two metals with the dielectric function $\epsilon_m$. From the other sides, the cavity is enclosed by a dielectric material to ensure an electrical isolation between the metal plates. To provide the possibility of detection of generated SPPs in the Kretschmann configuration, the whole structure is placed onto a prism with the dielectric function $\epsilon_p$, so that the substrate metal thickness, $h$, is of the order of the wavelength. We assume that the cavity interior is filled with the material of the dielectric function $\epsilon_h$ and contains identical spherical metal NPs randomly distributed with the volume fraction $f$.\\
Suppose now that at the moment of time $t=0$ one applies a constant voltage $V_0$ between the metal covers of the cavity. Then the polarization of NPs, ${\bf P}$, can be described in the framework of the harmonic oscillator model as follows \cite{Bordo16}
\begin{equation}\label{eq:oscillator}
\frac{d^2{\bf P}}{dt^2}+ \Gamma \frac{d{\bf P}}{dt}+\omega_0^2{\bf P} = a\left({\bf E}_0+\frac{4\pi}{3\epsilon_h}{\bf P}+{\bf E}_R\right),
\end{equation}
where $\omega_0$ is the frequency of the localized surface plasmon polariton (LSPP) supported by a nanoparticle, $\Gamma$ is the relaxation constant and the coefficient $a=(3/4\pi)f\epsilon_h\omega_0^2$ characterizes the coupling between the NPs and the electric field in the cavity. The cavity field is, in its turn, a sum of the Lorentz local field and the field scattered by the NPs and reflected back by the cavity walls, ${\bf E}_R$. We assume here that the effect of the walls static polarization due to the interaction with the NPs is already included in the constant field ${\bf E}_0$ created by the applied voltage.\\
In the linear regime, the solution of Eq. (\ref{eq:oscillator}) can be represented as a superposition of a constant contribution, dictated by the field ${\bf E}_0$, and the one oscillating with the frequency close to $\omega_0$, i.e. ${\bf P}(t)={\bf P}_0+{\bf P}_1(t)$, where ${\bf P}_0=(a/\bar{\omega}_0^2){\bf E}_0$ and $\bar{\omega}_0=\sqrt{\omega_0^2-(4\pi/3\epsilon_h)a}=\omega_0\sqrt{1-f}$ is the frequency of the LSPP renormalized because of the mutual interactions between the NPs.\\
The oscillating part of the polarization can be represented in the form 
\begin{equation}
{\bf P}_1(t)=\tilde{\bf P}_1(t)e^{-i\bar{\omega}_0t},
\end{equation}
where the amplitude $\tilde{\bf P}_1(t)$ varies in time much slower than $e^{-i\bar{\omega}_0t}$ and satisfies the equation
\begin{equation}\label{eq:P1}
\frac{d\tilde{\bf P}_1}{dt}+\frac{\Gamma}{2} \tilde{\bf P}_1 \approx i\beta \tilde{\bf E}_R
\end{equation}
with $\beta=a/(2\bar{\omega}_0)$ and the initial condition $\tilde{\bf P}_1(0)=-{\bf P}_0$. The slowly varying amplitude of the reflected field, $\tilde{\bf E}_R$, is in turn expressed in terms of $\tilde{\bf P}_1(t)$ through the approximate equation \cite{Bordo16}
\begin{equation}\label{eq:sca}
\tilde{{\bf E}}_R({\bf r},t)\approx\int\bar{{\bf F}}^R({\bf r},{\bf r}^{\prime};\bar{\omega}_0)\tilde{{\bf P}}_1({\bf r}^{\prime},t)d{\bf r}^{\prime},
\end{equation}
where $\bar{{\bf F}}^R({\bf r},{\bf r}^{\prime};\omega)$ is the reflected contribution to the field susceptibility tensor and the radius vectors ${\bf r}$ and ${\bf r}^{\prime}$ specify points in the cavity. The quantity $\bar{{\bf F}}^R({\bf r},{\bf r}^{\prime};\omega)$ relates the reflected electric field at the point ${\bf r}$ generated by a classical dipole, oscillating at frequency $\omega$, with the dipole moment itself, located at ${\bf r}^{\prime}$ \cite{Sipe84}. It can be obtained as a result of summation of multiple field reflections from the cavity walls. Its explicit form is known for a dipole between two parallel reflective surfaces \cite{Nha96}.\\
We assume that the dimensions of the cavity along the $x$ and $y$ axes are much larger than its height, i.e. $L_x,L_y\gg d$. In such a case, to a good approximation, the dipole emission in the cavity can be regarded as the one in an infinitely extended cavity \cite{Gerard09,Bordo12}. The corresponding field susceptibility tensor is given in the Supplemental Material \cite{SM}.\\
\emph{Criterion of generation}. -- Equation (\ref{eq:P1}) should be considered jointly with Eq. (\ref{eq:sca}) in order to analyze the stability of the cavity field. For a constant right-hand side part, it describes the transient oscillations of the polarization induced in the NPs decaying as $\exp(-\Gamma t/2)$. However, if Eq. (\ref{eq:sca}) establishes a positive feedback for the polarization oscillations, Eq. (\ref{eq:P1}) corresponds to 'negative damping' \cite{Jenkins13} which leads to the polarization self-oscillation and generation of the field in the cavity.\\
To investigate the field in the cavity, we expand it in the Fourier series over the intervals $-L_x/2\le x\le L_x/2$ and $-L_y/2\le y\le L_y/2$ as follows
\begin{eqnarray}
\tilde{{\bf E}}_R({\bf r},t)\nonumber\\
=\sum_{m,n=-\infty}^{\infty}{\bf e}^R_{mn}(z,t)\exp\left(i\frac{2\pi m}{L_x}x\right)\exp\left(i\frac{2\pi n}{L_y}y\right).
\end{eqnarray}
Similar expansions can be written for the quantities ${\bf E}_0({\bf r})$ and $\tilde{\bf P}_1({\bf r},t)$ with the coefficients ${\bf e}^0_{mn}(z)$ and ${\bf p}_{mn}(z,t)$, respectively. Substituting these expansions into Eqs. (\ref{eq:P1}) and (\ref{eq:sca}) and taking into account the inequalities $L_x,L_y\gg d$, one obtains the equations for the Fourier coefficients 
\begin{equation}\label{eq:p}
\frac{d{\bf p}_{mn}(z,t)}{dt}+\frac{\Gamma}{2} {\bf p}_{mn}(z,t) \approx i\beta {\bf e}^R_{mn}(z,t)
\end{equation}
and
\begin{equation}\label{eq:eR}
{\bf e}^R_{mn}(z,t)\approx\int_{-d/2}^{d/2}\bar{\mathcal{F}}^R(z,z^{\prime};\bar{\omega}_0,\kappa_{mn}){\bf p}_{mn}(z^{\prime},t)dz^{\prime},
\end{equation}
where $\bar{\mathcal{F}}^R(z,z^{\prime};\omega,\kappa)$ is the Fourier transform of the field susceptibility tensor \cite{SM} and 
\begin{equation}\label{eq:kappa}
\kappa_{mn}=2\pi\sqrt{\left(\frac{m}{L_x}\right)^2+\left(\frac{n}{L_y}\right)^2}
\end{equation}
is the absolute value of the wave vector of the field mode $(m,n)$.\\
Taking the Laplace transform of Eqs. (\ref{eq:p}) and (\ref{eq:eR}) in time and performing necessary integrations and summations one comes to the vector equation 
\begin{equation}\label{eq:vector}
\left[\hat{M}-\sigma(s)\hat{I}\right]\vec{A}(s)=\vec{B},
\end{equation}
where $\sigma(s)=(s+\Gamma/2)/i\beta$,
\begin{equation}
\vec{A}(s)=\left(\matrix{\mathcal{E}_{\kappa}^-(s)\cr
\mathcal{E}_{\kappa}^+(s)\cr
\mathcal{E}_z^-(s)\cr
\mathcal{E}_z^+(s)}\right),\quad
\vec{B}=\left(\matrix{\mathcal{P}_{\kappa}^{-}\cr
\mathcal{P}_{\kappa}^{+}\cr
\mathcal{P}_z^{-}\cr
\mathcal{P}_z^{+}}\right)
\end{equation}
with
\begin{equation}
\mathcal{E}_j^{\pm}(s)=\sum_k\int_{-d/2}^{d/2}\mathcal{H}_{jk}^{\pm }(z)\tilde{e}^R_k(z,s)dz,
\end{equation}
\begin{equation}
\mathcal{P}_j^{\pm}=-\frac{ia}{\beta\bar{\omega}_0^2}\sum_{kl}\int_{-d/2}^{d/2}\int_{-d/2}^{d/2}\mathcal{H}_{jk}^{\pm }(z)\mathcal{F}^R_{kl}(z,z^{\prime})e^0_l(z^{\prime})dzdz^{\prime}.
\end{equation}
Here $\hat{I}$ is the unit 4$\times$4 matrix, the explicit forms of the matrix $\hat{M}$ and the functions $\mathcal{H}_{jk}^{\pm }(z)$ are given in \cite{SM}, the tilde denotes the Laplace transform and we have omitted everywhere the subscripts $mn$ for the sake of brevity.\\
As it follows from Eq. (\ref{eq:vector}), the time evolution of the cavity field, which enters the vector $\vec{A}(s)$ by means of its Laplace transform $\tilde{\bf e}^R(s)$, is determined by the poles of $\vec{A}(s)$ or, equivalently, by the zeros $s_j$ of the determinant of the matrix $\hat{M}-\sigma(s)\hat{I}$. On the other hand, at these zeros the quantity $\sigma(s)$ gives the eigenvalues $\lambda_j$ ($j=1,...,4$) of the matrix $\hat{M}$ by definition. Therefore the poles of $\vec{A}(s)$ can be expressed as $s_j=-\Gamma/2+i\beta\lambda_j$. Consequently, if the matrix $\hat{M}$ has at least one eigenvalue with a negative imaginary part, $\text{Im}(\lambda_j)<0$, and 
\begin{equation}\label{eq:criterion}
-\text{Im}(\lambda_j)> \frac{\Gamma}{2\beta}=\frac{4\pi}{3}\frac{\Gamma}{f\epsilon_h\bar{\omega}_0},
\end{equation}
then the cavity field will increase with time as $\exp(gt)$ with the generation rate $g=-\Gamma/2-\beta\text{Im}(\lambda_j)$. In such a case the imaginary part of the corresponding pole will determine the frequency pulling effect for the frequency of generation,
\begin{equation}\label{eq:pulling}
\omega_g=\bar{\omega}_0-\beta\text{Re}(\lambda_j),
\end{equation}
which is known for lasers as well \cite{Pantell69}.\\
\emph{Steady-state operation.} -- Equation (\ref{eq:P1}) describes a linear regime of the NPs excitation when the cavity field is not too strong. If the generated field is very intensive, it should be corrected to take into account the nonlinear terms. The nonlinear optical response of metal NPs is manifested, in particular, as the saturation of absorption of metal-nanoparticle composite \cite{Plaksin08,Stepanov11}. It becomes essential when the exciting field intensity, $I$, is comparable with the saturation intensity, $I_s$. Then the nanocomposite absorption coefficient can be well described as $\alpha(I)=\alpha_0/(1+I/I_s)$ with $\alpha_0$ being the absorption coefficient in the linear regime. This effect can be introduced in Eq. (\ref{eq:P1}) by means of multiplying the coefficient $\beta$ by the factor $(1+I/I_s)^{-1}$. Then, considering the saturation regime where the cavity field amplitude $\tilde{\bf E}_R$ varies very slowly in time and performing the Laplace transform, one comes to the equation 
\begin{equation}\label{eq:vector1}
\left[(1+I/I_s)^{-1}\hat{M}-\sigma(s)\hat{I}\right]\vec{A}(s)=\vec{B}_1,
\end{equation}
instead of Eq. (\ref{eq:vector}). Here the vector $\vec{B}_1$ is determined by the value of the NPs polarization at some moment of time $t=t_1$ which corresponds to the saturation regime and we have neglected the time dependence of $I$ in the equation coefficient for the sake of simplicity. As before, the time evolution of the vector $\vec{A}$, and hence the cavity field, is determined by the zeros of the determinant of the matrix $(1+I/I_s)^{-1}\hat{M}-\sigma(s)\hat{I}$. Then the condition of the steady-state operation, $\text{Re}(s_j)=0$, gives the intensity of the generated field in the steady-state regime, $I_{ss}$:
\begin{equation}\label{eq:intensity}
\frac{I_{ss}}{I_s}=-\frac{2\beta}{\Gamma}\text{Im}(\lambda_j)-1
\end{equation}
with $\lambda_j$ being the eigenvalue of the matrix $\hat{M}$ which corresponds to generation. The generation frequency undergoes saturation as well:
\begin{equation}\label{eq:pulling_ss}
\omega_g=\bar{\omega}_0-\beta\text{Re}(\lambda_j)(1+I_{ss}/I_s)^{-1}.
\end{equation}
\emph{Numerical results.} -- We investigate the criterion of generation for a cavity between two silver films. One of them is deposited onto a prism surface ($\epsilon_p=1.45^2$) and has the thickness $h=100$ nm, whereas the other is much thicker than the wavelength of operation and is assumed to be semi-infinite in the $z$-direction. The cavity interior is filled with glass ($\epsilon_h=1.45^2$) and contains Ag NPs with the volume fraction $f$. The dielectric function of the Ag films is taken in the Drude model, $\epsilon_m(\omega)=\epsilon_{\infty}-\omega_p/[\omega(\omega+i\gamma)]$ with $\epsilon_{\infty}=5$, $\omega_p=14.0\times 10^{15}$ s$^{-1}$ and $\gamma=0.032\times 10^{15}$ s$^{-1}$ \cite{Shalaev10}. For silver NPs the relaxation constant can be written as $\Gamma=\gamma+bv_F/R$, where the Fermi velocity $v_F=1.4\times 10^4$ cm/s, $R$ is the NP radius and $b\approx 1$ \cite{Shalaev10}. For the given parameters and $f=0.01$, the LSPP wavelength is found as $\bar{\lambda}_0=2\pi c/\bar{\omega}_0=2\pi c\sqrt{\epsilon_{\infty}+2\epsilon_h}/(\omega_p\sqrt{1-f})\approx 410$ nm.\\
\begin{figure}
\includegraphics[width=\linewidth]{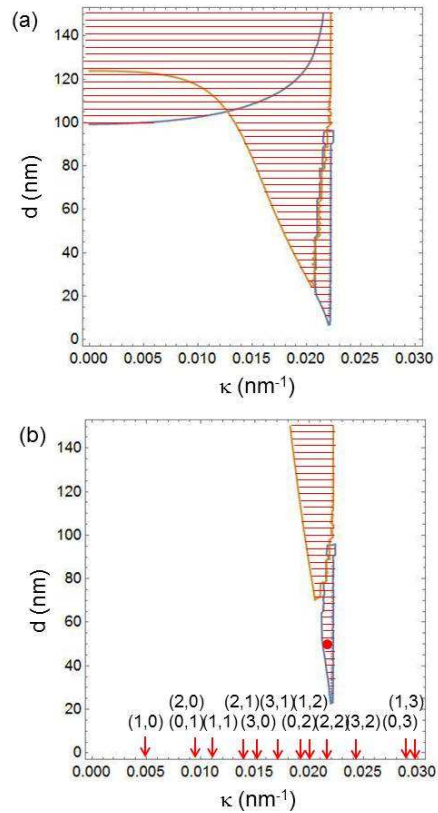}
\caption{\label{fig:criterion} (a) The contour plots $\text{Im}(\lambda_{1,2})=-\Gamma/(2\beta)$ in the plane $\kappa -d$ for $f=0.01$ and $R=10$ nm. The blue and brown lines correspond to different eigenvalues of the matrix $\hat{M}$. The area shaded in red is the region where the criterion of generation, Eq. (\ref{eq:criterion}), is fulfilled. (b) Same as $a$, but for $R=3$ nm. The red arrows show the values of $\kappa$ which correspond to different modes $(m,n)$, Eq. (\ref{eq:kappa}). The red dot specifies the set of parameters for which the numerical results are given in the text.}
\end{figure}
Figure \ref{fig:criterion} shows the contour plots in the parameter plane $\kappa-d$ which according to Eq. (\ref{eq:criterion}) determine the threshold for generation. The other eigenvalues of the matrix $\hat{M}$, $\lambda_{3,4}=0$, correspond to the polarization oscillations decaying with time as $\exp(-\Gamma t/2)$. The threshold is increased with the increase in $\Gamma$ (or, equivalently, with the decrease in the NP radius $R$) and with the decrease in $f$. Accordingly, the size of the region in the parameter space, where generation is possible, is reduced that can be used to ensure a single-mode operation.\\
The values of the modes wave numbers, Eq. (\ref{eq:kappa}), calculated for a cavity with $L_x=1300$ nm and $L_y=650$ nm are shown in Fig. \ref{fig:criterion}(b) by arrows. One can see that for the cavity thickness $d\approx 50$ nm the generation condition can be realized only for the mode $(2,2)$. However due to the degeneracy with respect to the signs of $m$ and $n$, the same condition is also fulfilled for the modes $(-2,2)$, $(2,-2)$ and $(-2,-2)$. These four modes propagate along the different directions specified by the azimuthal angles $\phi=\pm 63.4^{\circ},\pm 116.6^{\circ}$ in the $xy$ plane [see Fig. \ref{fig:sketch}(b)].\\
The initial intensity of the generated wave depends on the constant voltage, $V_0$, applied to the cavity. The calculation for $d=50$ nm and $\kappa_{22}=0.0216$ nm$^{-1}$gives for its mean amplitude at $t=0$ $\mid~e^R_z(0)~\mid \approx 0.30\times \mid~e^0_z(0)~\mid$, that corresponds to the initial wave intensity $I(0)\approx 1.1\times 10^8 \times \mid e^0_z(0)\mid^2$ in Gaussian units. If, for example, $V_0=1$ mV, then $I(0)\approx 5$ W/cm$^{2}$.\\ 
For the parameters given above, one finds the generation rate $g\approx 8.3\times 10^{13}$ s$^{-1}$. When the saturation comes into play, the intensity of the generated field in the steady-state regime can be obtained from Eq. (\ref{eq:intensity}) as $I_{ss}\approx 0.33I_s$. Taking into account the frequency pulling effect, Eq. (\ref{eq:pulling_ss}), one finds  the wavelength of generation $\lambda_g=407$ nm. From here one calculates the polar angle $\theta$ relative the $z$-axis, at which the generated wave can be detected in the Kretschmann configuration, as $\theta=\arcsin\{\kappa_{22}/[(2\pi/\lambda_g)\sqrt{\epsilon_p}]\}=74.8^{\circ}$.\\
For estimates we take the results of the self-consistent calculations of the saturable absorption in silica glass doped with Ag nanoparticles, which give $I_s\approx 100$ MW/cm$^2$ at $\lambda=430$ nm \cite{Kim10}. Despite a very high intensity of the generated wave ($I_{ss}\approx 33$ MW/cm$^2$), the corresponding consumed power is rather low: $P=I_{ss}(L_y/\cos\phi)d\approx 26$ mW.\\
\emph{Conclusion.} -- We have proposed and analyzed a self-excited generator of gap surface plasmon polaritons which is driven by a constant applied voltage. Its scheme is based on a plasmonic nanocavity doped with metal nanoparticles whose polarization undergoes a positive feedback from the reflective cavity walls. In contrast to spasers or plasmon lasers, such a generator does not exploit stimulated emission and does not require therefore powerful pumping, which is necessary to create a population inversion in a system with fast relaxation. The generation frequency is dictated by the LSPP frequency of the NPs and can be tuned by changing their metal composition, size, and shape \cite{Lee06}. The principal novelty of this approach, among other things, is the possibility to trigger the self-excitation (self-oscillation) process by applying a moderate electric field, that is a significant advantage for practical applications \cite{Duan03,Khurgin12}. 
\\
\emph{Acknowledgments.} -- The author is grateful to V.M.~Shalaev and A.S.~Lagutchev for fruitful discussions.


\begin{thebibliography}{99}
\bibitem{Atwater07} H.A.~Atwater, Sci. Am., April 2007, p. 56.
\bibitem{Ebbesen08} T.W.~Ebbesen, C.~Genet, and S.I.~Bozhevolnyi, Phys. Today, May 2008, p. 44. 
\bibitem{Barnes03} W.L.~Barnes, A.~Dereux, and T.W.~Ebbesen, Nature {\bf 424}, 824 (2003).
\bibitem{Schuller10} J.A.~Schuller, E.S.~Barnard, W.~Cai, Y.C.~Jun, J.S.~White, and M.L.~Brongersma, Nat. Mater. {\bf 9}, 193 (2010).
\bibitem{Zayats09} A.V.~Kabashin, P.~Evans, S.~Pastkovsky, W.~Hendren, G.A.~Wurtz, R.~Atkinson, R.~Pollard, V.A.~Podolskiy, and A.V.~Zayats, Nat. Mater. {\bf 8}, 867 (2009).
\bibitem{Novotny03} A.~Hartschuh, E.J.~S{\'a}nchez, X.S.~Xie, and L.~Novotny, Phys. Rev. Lett. {\bf 90}, 095503 (2003).
\bibitem{Sudarkin89} A.N.~Sudarkin and P.A.~Demkovich, Sov. Phys. - Tech. Phys. {\bf 34}, 764 (1989).
\bibitem{Pantell69} R.H.~Pantell and H.E.~Puthoff, {\it Fundamentals of Quantum Electronics} (John Wiley \& Sons, New York, 1969).
\bibitem{Stockman03} D.J.~Bergman and M.I.~Stockman, Phys. Rev. Lett. {\bf 90}, 027402 (2003).
\bibitem{Stockman11} M.I.~Stockman, Opt. Express {\bf 19}, 22029 (2011).
\bibitem{Ma13} R.-M.~Ma, R.F.~Oulton, V.J.~Sorger, and X.~Zhang, Laser Photonics Rev. {\bf 7}, 1 (2013).
\bibitem{Sommerfeld09} A.~Sommerfeld, Ann. Physik {\bf 28}, 665 (1909).
\bibitem{Sommerfeld64} A.~Sommerfeld, {\it Partial Differential Equations in Physics}, Lectures on Theoretical Physics, Vol. VI (Academic Press, New York,
London, 1964), Secs. 32, 33.
\bibitem{Bordo16} V.G.~Bordo, Phys. Rev. B {\bf 93}, 155421 (2016).
\bibitem{Jenkins13} A.~Jenkins, Phys. Rep. {\bf 525}, 167 (2013).
\bibitem{Sipe84} J.M.~Wylie and J.E.~Sipe, Phys. Rev. A {\bf 30}, 1185 (1984).
\bibitem{Nha96} H.~Nha and W.~Jhe, Phys. Rev. A {\bf 54}, 3505 (1996).
\bibitem{Gerard09} I.~Friedler, C.~Sauvan, J.P.~Hugonin, P.~Lalanne, J.~Claudon, and J.M.~G{\'e}rard, Opt. Express {\bf 17}, 2095 (2009).
\bibitem{Bordo12} V.~Bordo, J. Opt. Soc. Am. B {\bf 29}, 1799 (2012).
\bibitem{SM} See Supplemental Material for the field susceptibility tensor and the matrix $\hat{M}$.
\bibitem{Plaksin08} O.~Plaksin, Y.~Takeda, H.~Amekura, N.~Kishimoto, and S.~Plaksin, J. Appl. Phys. {\bf 103}, 114302 (2008).
\bibitem{Stepanov11} A.L.~Stepanov, Rev. Adv. Mater. Sci. {\bf 27}, 115 (2011).
\bibitem{Shalaev10} W.~Cai and V.~Shalaev, {\it Optical Metamaterials} (Springer, New York, 2010).
\bibitem{Kim10} K.-H.~Kim, A.~Husakou, and J.~Herrmann, Opt. Express. {\bf 18}, 21918 (2010).
\bibitem{Lee06} K.-S.~Lee and M.A.~El-Sayed, J. Phys. Chem. B {\bf 110}, 19220 (2006).
\bibitem{Duan03} X.~Duan, Y.~Huang, R.~Agarwal, and C.M.~Lieber, Nature {\bf 421}, 241 (2003).
\bibitem{Khurgin12} J.B.~Khurgin and G.~Sun, Opt. Express {\bf 20}, 15309 (2012).






\end{thebibliography}
\end{document}